\documentclass[12pt,report]{elsart}

\usepackage{epsfig,graphics,colordvi}
\usepackage{amsmath}
\usepackage{amssymb}
\usepackage{color}

\begin{document}
\newcommand{\pslash}{\slash p}
\newcommand{\barqslash}{\FMslash {\bar q}}
\newcommand{\wslash}{\FMslash w}
\newcommand{\qslash}{\FMslash q}
\newcommand{\Qslash}{\FMslash Q}
\newcommand{\Pslash}{\FMslash P}
\newcommand{\notsubset}{\;\FMslash \subset \;}
\begin{frontmatter}
\title{Compton scattering from chiral dynamics with unitarity and causality}
\author[GSI,ITEP]{A.M. Gasparyan,}
\author[GSI]{M.F.M. Lutz,}
\author[Pavia]{B. Pasquini}
\address[GSI]{GSI Helmholtzzentrum f\"ur Schwerionenforschung GmbH,\\
Planckstra\ss e 1, 64291 Darmstadt, Germany}
\address[ITEP]{SSC RF ITEP, Bolshaya Cheremushkinskaya 25,\\ 117218 Moscow, Russia}
\address[Pavia]{Dipartimento di Fisica Nucleare e Teorica,\\ Universit`a degli Studi di Pavia and INFN, Sezione di Pavia, Pavia, Italy}
\begin{abstract}
Proton Compton scattering is analyzed with the chiral Lagrangian.
Partial-wave amplitudes are obtained by an analytic extrapolation of
subthreshold reaction amplitudes computed in chiral perturbation theory, where the constraints set by
electro\-magnetic-gauge invariance, causality and unitarity are used to stabilize the extrapolation.
We present and discuss predictions for various spin observables and
polarizabilties of the proton. While for the transition polarizabilities $\gamma_{E1M2}, \gamma_{M1E2}$
we recover the results of strict chiral perturbation theory, for the diagonal $\gamma_{E1E1}, \gamma_{M1M1}$ elements
we find significant effects from rescattering.

\end{abstract}
\end{frontmatter}

\section{Introduction}

Nucleon Compton scattering serves as an important test for Chiral Perturbation Theory ($\chi $PT). One of the reasons
is that for such a process the leading order loop contributions do not depend on unknown parameters, when calculated
in the heavy baryon formalism \cite{Bernard:1995cj}. The same is true for some of the higher order terms
\cite{Gellas:2000mx,VijayaKumar:2000pv}. The application of strict $\chi $PT is however limited to
the near threshold region. A method to extrapolate $\chi $PT results beyond the threshold region
using analyticity and unitarity constraints was proposed recently in \cite{Gasparyan:2010xz,Danilkin:2010xd,Gasparyan:2010fb}. The results for the
differential cross section and beam asymmetry for the proton Compton scattering from threshold up to
$\sqrt{s}\approx 1300$ MeV are in agreement with empirical data even though all the parameters were
determined from $\pi N$ elastic scattering and pion photoproduction. The spin-independent polarizabilities of the
proton in this approach do not differ from  $\chi $PT prediction by construction \cite{Gasparyan:2010xz}
and are equal to $\alpha_p=13.0\cdot 10^{-4}\,\rm{fm}^3$, $\beta_p=1.3\cdot 10^{-4}\,\rm{fm}^3$ in the heavy baryon
formalism at the order $Q^3$. The spin-dependent polarizabilities and further polarization observables such as double
polarization observables  $\Sigma_{2x}$ and $\Sigma_{2z}$  were not discussed in \cite{Gasparyan:2010xz}.
Double polarization observables are of particular interest since they are best suited
to access the spin-dependent polarizabilities. Measurements are under way at MAMI~\cite{MAMIA2-11/09} in the energy range above pion-production threshold, and experiments at lower energies are also planned  at the HI$\gamma$S facility~\cite{Weller:2009zza}.

It is the purpose of this work to derive such quantities in our unitary and effective field theory and compare with
previous phenomenological approaches based on fixed-$t$ dispersion relations (DRs)
\cite{Drechsel:1999rf,Pasquini:2007hf}.

\section{Chiral symmetry, analyticity and unitarity}

Our approach is based on the chiral Lagrangian involving pion, nucleon and photon fields \cite{Fettes:1998ud,Bernard:2007zu}.
The terms relevant at the order $Q^3$ for pion elastic scattering,
pion photoproduction and nucleon Compton scattering are taken into account
\footnote{Note a typo in Eq.~(1) of \cite{Gasparyan:2010xz}. }
\allowdisplaybreaks[1]
\begin{eqnarray}
\mathcal{L}_{int}&=&
-\frac{1}{4\,f^2}\,\bar{N}\,\gamma^{\mu}\,\big( \vec{\tau} \cdot
\big(\vec{\pi}\times
(\partial_\mu\vec{\pi})\big)\big) \,N +
\frac{g_A}{2\,f} \,\bar{N}\,\gamma_5\,\gamma^{\mu} \,\big(
\vec{\tau}\cdot (\partial_{\mu}\vec{\pi} )\big) \,N
\nonumber \\
&-&e\,\Big\{ \big(\vec{\pi}\times(\partial_{\mu}\vec{\pi}) \big)_3
+ \bar{N}\,\gamma_\mu\, \frac{1+\tau_3}{2} \,N
- \frac{g_A}{2\,f}
\,\bar{N}\,\gamma_5\,\gamma_{\mu}\,\big(\vec\tau\times\vec{\pi}\big)_3\,N \Big\}
\,A^\mu
\nonumber\\
&-&\frac{e}{4\,m_N}\,\bar{N}\,\sigma_{\mu\nu}\,\frac{\kappa_s+\kappa_v\,\tau_3}{2}\,N\,F^{\mu\nu}+
\frac{e^2}{32\pi^2
f}\,\epsilon^{\mu\nu\alpha\beta}\,\pi_3\,F_{\mu\nu}\,F_{\alpha\beta}
\nonumber\\
&-&\frac{2\,c_1}{f^2}\,m_\pi^2\, \bar{N}\,( \vec{\pi}\cdot\vec{\pi})\,N -
\frac{c_2}{2\,f^2\,m_N^2}\,\Big\{\bar{N}\,(
\partial_{\mu}\,\vec{\pi})\cdot (\partial_{\nu}\vec{\pi})\,(\partial^\mu
\partial^\nu N )+\rm{h.c.}\Big\}
\nonumber \\
&+& \frac{c_3}{f^2}\,\bar{N} \,(\partial_{\mu}\,\vec{\pi} )
\cdot (\partial^{\mu}\vec{\pi})\,N
-\frac{c_4}{2\,f^2}\,\bar{N}\,\sigma^{\mu\nu}\,\big(\vec{\tau} \cdot
\big((\partial_{\mu}\vec{\pi})\times
(\partial_{\nu}\vec{\pi})\big)\big)\,N
\nonumber\\
&-&i\,\frac{d_1+d_2}{f^2\,m_N}\,
\bar{N}\,\big(\vec\tau\cdot \big((\partial_\mu \vec \pi )\times
(\partial_\nu\partial_\mu \vec\pi) \big)\big) \, (\partial^\nu N) +
\rm{h.c.}
\nonumber \\
&+&\frac{i\,d_3}{f^2\,m_N^3}\,
\bar{N}\,\big(\vec \tau \cdot \big( (\partial_\mu\vec\pi )\times
(\partial_\nu \partial_\lambda\vec\pi )\big)\big)\,
(\partial^\nu\partial^\mu\partial^\lambda N)
+\mbox{h.c.}
\nonumber\\
&-&2\,i\,\frac{m_\pi^2\,d_5}{f^2\,m_N}\,\bar{N}\,\big(\vec{\tau} \cdot
\big(\vec{\pi}\times
(\partial_\mu\vec{\pi}) \big)\big)\,( \partial ^\mu N) +\rm{h.c.}
\nonumber\\
&-&\frac{i \,e }{f\,m_N }\, \epsilon^{\mu\nu\alpha\beta}\,\bar{N}\,\big(
d_8\,
(\partial_\alpha \,\pi_3) +d_9\,\big(\vec\tau \cdot (\partial_\alpha
\vec \pi)\big) \big)\, (\partial_\beta\,N)\, F_{\mu\nu}+\mbox{h.c.}
\nonumber \\
&+&i\,\frac{d_{14}-d_{15}}{2\,f^2\,m_N}\,
\bar{N}\,\sigma^{\mu\nu}\,\big((\partial_\nu\vec\pi )\cdot
(\partial_\mu\partial_\lambda\vec\pi ) \big)\,(\partial^\lambda N)
+\mbox{h.c.}
\nonumber \\
&-&\frac{m_\pi^2\,d_{18}}{f} \,\bar{N}\,\gamma_5\,\gamma^{\mu} \, \big(
\vec{\tau}\cdot (\partial_{\mu}\vec{\pi})\big) \, N
-\frac{e\,m_\pi^2\,d_{18}}{f} \,\bar{N}\,\gamma_5\,\gamma^{\mu} \, \big(
\vec{\tau}\times \vec{\pi}\big)_3 \, N\,A_\mu
\nonumber \\
&+&\frac{e \,(d_{22}-2\,d_{21})}{2\,f}\,\bar{N}\,\gamma_5\,\gamma^\mu\,
\big(\vec\tau\times\partial^\nu \,\vec \pi\big)_3 \,N\, F_{\mu\nu}
\nonumber \\
&+&\frac{e\, d_{20}}{2 \,f\,m_N^2}\,\bar{N}\,\gamma_5\,\gamma^\mu\,
\big(\vec\tau \times (\partial_\lambda \,\vec \pi)\big)_3\,
(\partial^\nu\partial^\lambda N)\, F_{\mu\nu}+\mbox{h.c.} \,.
\label{Lagrangian}
\end{eqnarray}
A strict chiral expansion of the various scattering amplitudes to the order
$Q^3$ includes tree-level graphs, loop diagrams, and counter terms. The latter depend on a few low energy
constants ``$c_i$' and ``$d_i$'', which are to be adjusted to the empirical data set.
Our goal is to improve the convergence of strict $\chi $PT by an analytic continuation
of the subthreshold amplitudes. As compared to strict $\chi $PT the applicability domain is extended towards
higher energies beyond the threshold region \cite{Gasparyan:2010xz}. The extrapolation of the amplitudes is performed
insisting on basic principles of analyticity and unitarity. We seek to separate left- and right-hand singularities of
suitably chosen partial-wave amplitudes with angular
momentum $J$, parity $P$ and channel quantum numbers $a,b$ (in our case $\pi N$ and $\gamma N$ channels). The
separation  is achieved by means of the partial-wave dispersion relation
\begin{eqnarray}
&& T_{ab}^{JP}(\sqrt{s}\,)=U_{ab}^{JP}(\sqrt{s}\,)
\nonumber\\
&& \qquad +\sum_{c,d} \int_{\mu_{\rm
thr}}^{\infty}\frac{dw}{\pi}\frac{\sqrt{s}-\mu_M}{w-\mu_M}
\frac{T_{ac}^{*,JP}(w)\,\rho^{JP}_{cd}(w)\,T_{db}^{JP}(w)}{w-\sqrt{s}-i\epsilon}\,,
\label{disrel}
\end{eqnarray}
where the generalized potential, $U_{ab}^{(JP)}(\sqrt{s}\,)$, is the
part of the amplitude that contains left-hand cuts only. 
The phase-space
matrix $\rho^{JP}_{cd}(w)$ reflects our particular convention for the
partial-wave amplitudes, that are free of kinematic constraints. The dispersion relation (\ref{disrel})
relies on the fact that the discontinuity along the right-hand cut is given by the unitarity condition
\begin{eqnarray}
 \nonumber \Delta T_{ab}^{(JP)}(\sqrt{s})&=&\frac{1}{2\,i}\left(T_{ab}^{(JP)}(\sqrt{s}+i\epsilon)-
T_{ab}^{(JP)}(\sqrt{s}-i\epsilon)\right)\\
&=&\sum_{c,d}\,T_{ac}^{(JP)}(\sqrt{s}+i\epsilon)\,\rho^{(JP)}_{cd}(\sqrt{s}\,)\,T_{db}^{(JP)}(\sqrt{s}-i\epsilon).
\label{discontinuity}
\end{eqnarray}
It is emphasized that the separation (\ref{disrel}) is gauge invariant. This follows since both contributions
in (\ref{disrel}) are strictly on-shell and characterized by distinct analytic properties. The relevance of the 
subtraction scale $\mu_M$ in (\ref{disrel}) will be discussed below. 

In the present approach we take into account only the $\pi N$ intermediate states in (\ref{disrel}) (channels $c,d$ in
the sum). The higher mass intermediate states are then effectively included in the potential $U_{ab}^{(JP)}(\sqrt{s})$.
We also neglect the $\gamma N$ intermediate state, since the effects thereof are suppressed by the square of the
electromagnetic charge.

For a given approximative generalized potential the nonlinear integral
equation (\ref{disrel}) yields scattering amplitudes, that obey the constraints set by the unitarity condition.
On the other hand the crossing symmetry constraint is not automatically satisfied. If crossing symmetry would be
implemented exactly, a crossed reaction can be computed from the direct reaction amplitude. Necessarily, the
crossed reaction would be determined by the direct reaction amplitude evaluated at energies below the $s$-channel
threshold. Now, imagine that in a first step we have some means to approximate the generalized potential systematically
and very accurately at energies above the $s$-channel threshold only. We may use this potential and determine via
(\ref{disrel}) the scattering amplitude in the physical region. In this case we are not able to compute the crossed
reaction, since our approximation for the generalized potential is valid above threshold only, by assumption.
The crossed reaction may be computed by setting up the analogous nonlinear integral
equation (\ref{disrel}) for the crossed reaction. It involves potentials distinct from the potentials of the
direct reaction. But where is the crossing symmetry constraint? Our argument illustrates that crossing symmetry is a
constraint that affects dominantly amplitudes at subthreshold energies. In a second step we now
imagine that we have some means to approximate the generalized potential systematically at energies
$\sqrt{s}\geq  \mu_M $, i.e. we include a small subthreshold region. In this case crossing symmetry does provide
a constraint. The crossed reaction amplitude can be computed from the direct amplitudes, however, only  in  a specific
subthreshold region. The coincidence of the crossing transformed amplitudes and the amplitudes from the crossed reaction
in that specific subthreshold region defines the desired constraint. Note that a nonempty coincidence region requires
$\mu_M$ to be distinct from both $s$- and $u$-channel unitarity branch points. Our particular choice
$\mu_M=m_N$ implies a significant size of the coincidence region.

The crucial assumption of our approach is the perturbative nature of the reaction amplitudes at subthreshold energies,
where the crossing symmetry constraints are relevant. To this extent we satisfy crossing symmetry in a
perturbative manner. In order to arrive at even improved coincidence properties we insist on a subtraction
at $\sqrt{s}=\mu_M$ in (\ref{disrel}). Since the generalized potential of the direct and
crossed reactions are evaluated in a perturbative manner, the crossing symmetry properties of perturbation theory are
transported into our approach.

We recall the specifics of how to approximate the generalized potential \cite{Gasparyan:2010xz}.
In an initial step the various reaction amplitudes are evaluated in strict $\chi$PT to a given order, in our case $Q^3$.
From this result we may extract the generalized potential accurate to the same chiral order. In a second step
we apply conformal mapping techniques to extrapolate the potential into a larger energy domain $\Omega$.
This is done by means of a transformation $\xi(\sqrt{s})$ that maps $\Omega$ onto the unit circle. The domain $\Omega$
includes energies significantly above the $s$-channel threshold, but excludes energies below the $u$-channel unitarity
threshold necessarily. While it is possible to extrapolate the potential towards higher energies, it is not 
straightforward to extrapolate it to energies below the $u$-channel unitarity threshold. This is a consequence of the analytic
structure of the potential. Since the potential is characterized by left-hand cut structures it is an analytic function
for energies larger than the $s$-channel unitarity threshold. The presence of all left-hand cut structures imply that the
potential is not analytic for energies sufficiently below the $s$-channel unitarity threshold.

All together the generalized potential can be represented
as a series in $\xi$ convergent in the domain $\Omega$ with
\begin{eqnarray}
&& U(\sqrt{s}\,)=U_{\rm inside}(\sqrt{s}\,)+U_{\rm outside}(\sqrt{s}\,)\,,
\label{expansion} \\
&& U_{\rm outside}(\sqrt{s}\,)=\sum\limits_{k=0}^\infty {U_k}\,\big[\xi(\sqrt{s}\,)\big]^k\,, \qquad
U_k=\frac{d^k U_{\rm outside}(\xi^{-1}(\xi))}{k!\,d\xi^k}\Big|_{\xi=0}\,,
\nonumber
\end{eqnarray}
where we allow for an explicit treatment of cut structures that are inside the domain $\Omega$.
These are contained in $U_{\rm inside}(\sqrt{s}\,)$ and correspond typically to contributions from tree-level diagrams.
The coefficients $U_k$ are uniquely determined from the Taylor expansion of the generalized potential in the vicinity
of the $s$-channel threshold. To the order we are working at, we truncate the series in (\ref{expansion}) at the zeroth
or the first term depending on the channel and the partial wave.

We close our brief primer on the essential ingredients of the novel scheme introduced in \cite{Gasparyan:2010xz}
with a discussion of so called CDD poles \cite{Johnson:1979jy,Castillejo:1955ed}. In general there is no guarantee
that a solution of the non-linear integral equation (\ref{disrel}) exists and/or is unique. Different solutions
may be characterized by the presence of distinct CDD poles in the $N/D$ ansatz that solves (\ref{disrel}).
We need a condition which singles out the unique and physical solution we are interested in. We require that
the non-linear equation can be solved in perturbation theory at least close to the matching scale $\mu_M$. To be
specific, the reaction amplitude must be accessible via an iterative solution of (\ref{disrel}) in a small
domain around $\mu_M$. Two remarks are here in order. First, such a condition implies that the reaction
amplitudes are consistent with the crossing symmetry constraint, as discussed in detail above. Second,
this condition excludes the presence of a CDD pole. In a case where this iterative requirement cannot be met we admit the
presence of CDD poles \cite{Johnson:1979jy,Castillejo:1955ed}. As shown in  \cite{Gasparyan:2010xz} the
perturbative nature of the reaction amplitudes can be restored by a suitable correlation of the CDD pole parameters.
There is no problem with the approximate crossing symmetry, even though a solution with a CDD pole
can not be computed iteratively close to $\mu_M$.

We need to include CDD poles in the $P_{11}$ and $P_{33}$ partial waves corresponding to Delta and Roper resonances.
These CDD poles represent an infinite and correlated summation of higher order counter terms of the chiral
Lagrangian (\ref{Lagrangian}). For more technical details we refer to \cite{Gasparyan:2010xz}. Note that our
approach is at variance with other approaches where the isobar field is introduced as an explicit degree of freedom
in the chiral Lagrangian~\cite{Hildebrandt:2003fm,Hildebrandt:2003md,Lensky:2009uv}. We do not argue against
an explicit isobar field in the chiral Lagrangian. In fact a possible extension of our
approach would include an explicit isobar field in the chiral Lagrangian. However, so far all our results indicate
that the most important ingredients for successful descriptions and predictions of pion and photon induced reactions
off the nucleon based on the chiral Lagrangian are unitarity and analyticity.

\section{Spin polarizabilities of the proton}

The spin structure of the nucleon is described by the effective interaction
of the third order
\begin{eqnarray}
&&{H}_{\rm{eff}} =  - 4 \pi\, \big[
{\textstyle\frac{1}{2}}\,\gamma_{E1E1}
\,\vec{\sigma}\cdot(\vec{E}\times\dot{\vec{E}})
+ {\textstyle\frac{1}{2}}\,\gamma_{M1M1}\,
\vec{\sigma}\cdot(\vec{H}\times\dot{\vec{H}})  \nonumber \\
&&\;
- \,{\textstyle\frac{1}{2}}\,\gamma_{M1E2}\, (\vec \nabla \cdot  \vec{\sigma} \,E_{i}
                   + \nabla_i   \vec E\, \cdot \vec{\sigma} )\,H_i
+{\textstyle\frac{1}{2}}\, \gamma_{E1M2}\,(\vec \nabla \cdot  \vec{\sigma} \,H_{i}
               + \nabla_i \vec H \cdot \vec{\sigma}  )\,E_i  \big],\,
\end{eqnarray}
where the four spin polarizabilities $\gamma_{E1E1}$,$\gamma_{M1M1}$, $\gamma_{M1E2}$, $\gamma_{E1M2}$ are related
to the multipole expansion \cite{Babusci:1998ww}. Spin polarizabilities can be also defined in terms of threshold
values of invariant scattering amplitudes. Here we apply a set of amplitudes, $F_{1-3}^\pm(\sqrt{s},t)$, introduced
in  \cite{Gasparyan:2010xz}. They  are particularly well suited for the construction of partial-wave amplitudes,
$T^J_{\pm,ab}(\sqrt{s}\,)$, that are free from kinematical constraints and therefore useful in dispersion-integral
representations like (\ref{disrel}). Owing to the MacDowel relations
\begin{eqnarray}
F_i^{(-)}(+\sqrt{s}, \,t) = F_i^{(+)}(-\sqrt{s}, \,t)\,,
\label{}
\end{eqnarray}
there are in fact only three independent amplitudes \cite{MacDowell:1959zza,Gasparyan:2010xz}. Similar
MacDowell relations hold for the partial-wave amplitudes
\begin{eqnarray}
T_{-,ab}^J(+\sqrt{s}\,)= \left\{
\begin{array}{ll}
+ T_{+,ab}^J(-\sqrt{s}\,) \qquad {\rm for} & \quad a=b \\
- T_{+,ab}^J(-\sqrt{s}\,) \qquad {\rm for} & \quad a\neq b
\end{array}
\right.
\,,
\label{MacDowell-Compton}
\end{eqnarray}
where $a,b=1,2$ run over the two possible photon spins. Note that the more conventional multipole amplitudes
introduced in \cite{Babusci:1998ww} do not satisfy the MacDowell relations (\ref{MacDowell-Compton}). The Compton
scattering tensor $T_{\mu\nu}(\bar q, q;w)$ reads
\begin{eqnarray}
&&T_{\mu\nu}(\bar q, q;w)= \sum_{i=1}^3\, \Big( F^+_{i}(\sqrt{s},t)\,L^{(i,+)}_{\mu \nu}(\bar q,q,w)
+F^-_{i}(\sqrt{s},t)\,L^{(i,-)}_{\mu \nu}(\bar q,q,w)\Big) \,,
\nonumber\\
&&  L^{(1,\pm )}_{\mu \nu} = \gamma_\nu \,\qslash \,
\Big( \frac{1}{2}\pm \frac{\wslash }{2\,\sqrt{s}} \Big)\,\barqslash\,\gamma_\mu \,, \qquad
L^{(2,\pm)}_{\mu \nu} =\gamma_\mu \,\barqslash \,\Big( \frac{1}{2}\pm \frac{\wslash }{2\,\sqrt{s}} \Big)\,\qslash\,\gamma_\nu  \,,
\nonumber\\
&& L^{(3,\pm)}_{\mu \nu} = \Big( \frac{1}{2}\pm \frac{\wslash }{2\,\sqrt{s}} \Big)\,
\Big((w\cdot \bar{q})\,g_{\mu\alpha}-w_\mu\,\bar{q}_\alpha\Big)\,
\Big(q_\alpha\gamma_\nu-\qslash \,g_{\alpha\nu}\Big)
\nonumber\\
&& \qquad \;\;\,+ \,\Big((w\cdot q)\,g_{\nu\alpha}-w_\nu\,q_\alpha\Big)\,
\Big(\bar q_\alpha\gamma_\mu-\barqslash \,g_{\alpha\mu}\Big)\,\Big( \frac{1}{2}\pm \frac{\wslash }{2\,\sqrt{s}} \Big) \,,
\label{def-Fpm123}
\end{eqnarray}
where $q$ and $\bar q$ are the 4-momenta of the initial and final photons respectively, and
$w$ is the total 4-momentum of the photon-nucleon system.

The spin polarizabilities are most economically identified in terms of threshold values of
amplitudes, $A_{1-6}(\nu,t)$, that are even under a crossing transformation \cite{L'vov:1996xd}. We derive
their relation with the MacDowel amplitudes
\allowdisplaybreaks[1]
\begin{eqnarray}
&& A_{i}(-\nu,\,t) =+  A_{i}(+\nu,\,t) \,, \qquad \qquad \nu=\frac{s-u}{4\,m_p} \,,
\nonumber\\
&& A_1=\frac{1}{2} \,(F_1^+ + F_2^+)+\frac{m_p^2 - 2 \,m_p \sqrt{s}-3\,s - 2\,\nu\,\sqrt{s}}{4\,\sqrt{s}}
F_3^++(\sqrt{s}\leftrightarrow-\sqrt{s}\,)\,,
\nonumber\\
&&A_2=\frac{1}{2} \, (F_1^+ + F_2^+) + \frac{m_p^2 +s-2\,\nu\,\sqrt{s}}{4\, \sqrt{s}}\,F_3^+
+(\sqrt{s}\leftrightarrow-\sqrt{s}\,)\,,
\nonumber\\
&&A_3=\frac{m_p^2}{2\, \nu \sqrt{s}}\, (F_2^+ - F_1^+ - F_3^+ \nu)
+(\sqrt{s}\leftrightarrow-\sqrt{s}\,)\,,
\nonumber\\
&&A_4=\frac{m_p^2}{2\, \nu \sqrt{s}}\,(F_2^+ - F_1^+)+(\sqrt{s}\leftrightarrow-\sqrt{s}\,)\,,
\nonumber\\
&&A_5=\frac{m_p}{2\,\nu} \,(F_2^+ - F_1^+) +
\frac{(s-m_p^2)(\sqrt{s}-m_p)-2\,m_p\,\nu\,\sqrt{s}}{4\,\nu\,\sqrt{s}}F_3^+\,
\nonumber\\
&& \quad\,+(\sqrt{s}\leftrightarrow-\sqrt{s}\,)\,,
\nonumber\\
&& A_6=\frac{m_p^2 + s}{4\,\nu \sqrt{s}}\,(F_1^+ -F_2^+)  -\frac{m_p}{\sqrt{s}}\, F_1^+
-\frac{m_p}{2}\, F_3^+ +(\sqrt{s}\leftrightarrow-\sqrt{s}\,)\,.
\label{AviaF}
\end{eqnarray}
The amplitudes $A_i(\nu,t)$ are free of kinematic singularities and constraints and are even functions
of $\nu$, which is a consequence of crossing symmetry. The spin polarizabilities are expressed in
terms of the coefficients $a_{i}$, that are defined by threshold values of Born-subtracted amplitudes, $A_i(\nu,t)$.
It holds
\begin{eqnarray}
&& a_i = A_i(0, 0) - A_i^{Born}(0, 0)\,,
\nonumber\\
&& \gamma_{\,E1\,E1}=\frac{a_2-a_4+2 a_5+a_6}{8\pi\, m_p}\,\,, \qquad \quad \gamma_{E1M2}=\frac{a_2-a_4-a_6}{8\pi \,m_p}\,,
\nonumber\\
&& \gamma_{M1M1}=-\frac{a_2+a_4+2a_5-a_6}{8\pi \,m_p}\, ,\qquad  \,
\gamma_{M1E2}=-\frac{a_2+a_4+a_6}{8\pi \,m_p}\,.
\end{eqnarray}

While our approach respects the unitarity and causality constraints exactly, crossing symmetry is conserved
at an approximate level only.  As a consequence the amplitudes $A_i(\nu,t)$ will not be even functions in $\nu$.
Depending on the detailed form of the crossing violations some amplitudes could even be singular in the
limit $\nu \to 0$. This is not the case in our approach and therefore it is possible to identify the
spin polarizabilities in an unambiguous manner.

From (\ref{AviaF}) it follows that the regularity of the amplitude $A_i(\nu,t)$ for $i=3,4,5,6$ requests the relations
\begin{eqnarray}
F_1^\pm(m_p,0)=F_2^\pm(m_p,0) \,,
\end{eqnarray}
for the Born subtracted amplitudes. This is realized in our approach due to the matching condition at
$\sqrt{s}=m_p$  in (\ref{disrel}) that insures that the partial-wave amplitudes at this point coincide with
their $\chi$PT values \cite{Gasparyan:2010xz}.

In the  Appendix~\ref{PWamplitudes} we express the invariant amplitude $F^\pm_i(\sqrt{s}, t)$ in terms of
the partial-wave amplitudes  $T_{\pm,ab}^J(\sqrt{s}\,)$. As an application we find
\begin{eqnarray}
\gamma_{\,E1\,E1}&=&\frac{1}{16\pi \,m_p^3}\,\frac{d}{d\,\sqrt{s}\,} \Big[
\frac{4}{3}\, \Delta T^{\frac32}_{-,22}(m_p) -\Delta T^{\frac12}_{+,11}(m_p) \Big]
\nonumber\\
&+&\frac{1}{32\pi \,m_p^4} \,\Big[ \frac{4}{\sqrt{3 }}\, \Delta T^{\frac32}_{-,21}(m_p)
- \Delta T^{\frac12}_{+,11}(m_p)\Big] \, ,
\nonumber\\
\gamma_{M1M1}&=&\frac{1}{16\pi \,m_p^3}\,\frac{d}{d\,\sqrt{s}\,} \Big[
\frac{4}{3}\, \Delta T^{\frac32}_{+,22}(m_p) - \Delta T^{\frac12}_{-,11}(m_p) \Big]
\nonumber\\
&+&\frac{1}{32\pi \,m_p^4} \,\Big[ \frac{4}{\sqrt{3 }}\, \Delta T^{\frac32}_{+,21}(m_p)
- \Delta T^{\frac12}_{-,11}(m_p)\Big] \, ,
\nonumber\\
\gamma_{\,E1M2}&=&\frac{3}{32\pi m_p^4} \,\Big[\Delta T^{\frac12}_{+,11}(m_p)
-\frac{4}{ \sqrt{3}}\, \Delta T^{\frac32}_{-,21}(m_p) \Big] \, ,
\nonumber\\
\gamma_{\,M1E2}&=&\frac{3}{32\pi m_p^4} \,\Big[\Delta T^{\frac12}_{-,11}(m_p)
-\frac{4}{ \sqrt{3 }}\, \Delta T^{\frac32}_{+,21}(m_p) \Big] \, ,
\label{polarizabilities:PW}
\end{eqnarray}
where $\Delta T^{J}_{\pm,ab}(\sqrt{s}\,)$ stays for the non-Born part of the partial-wave amplitude.
Our result (\ref{polarizabilities:PW}) has an interesting consequence. While for the  polarizabilities
$\gamma_{E1M2}, \gamma_{M1E2}$ we recover the results of strict $\chi$PT, for $\gamma_{E1E1}, \gamma_{M1M1}$
rescattering effects are involved in terms of threshold derivatives of partial-wave amplitudes.

At present there are empirical constraints on two combinations of the
polarizabilities \cite{Ahrens:2001qt,Dutz:2003mm,Pasquini:2010zr,Schumacher:2005an,Arndt:2002xv}.
The forward ($\gamma_0$) and backward ($\gamma_\pi$) polarizabilities are\begin{eqnarray}
\gamma_{0}&=&-\gamma_{E1E1}-\gamma_{M1M1}-\gamma_{E1M2}-\gamma_{M1E2}  \nonumber\\
&=& (-1.01\pm0.08\pm0.13) \,10^{-4}{\rm fm}^4\,,\\
\nonumber\\
\gamma_{\pi}&=&-\gamma_{E1E1}+\gamma_{M1M1}-\gamma_{E1M2}+\gamma_{M1E2}
\nonumber\\
&=&(8.0\pm1.8 )\,10^{-4}{\rm fm}^4 \,.
\end{eqnarray}
The comparison of our results for the spin polarizabilities with other theoretical approaches is presented
in Table~\ref{table:polarizabilities}.
\begin{table*}[b]
\parbox{6.5cm}{
\begin{center}
\begin{tabular}{|c|c|c|c|c|}
\hline
&$\chi$PT, $Q^3$ \cite{VijayaKumar:2000pv} &$\chi$PT, $Q^4$ \cite{VijayaKumar:2000pv}&DR \cite{Holstein:1999uu,Drechsel:2002ar}&this work\\
\hline
$\gamma_{E1E1}$&$-5.93$&$-1.41$&$-4.3$&$-3.68$\\
\hline
$\gamma_{M1M1}$&$-1.19$&$3.38$&$2.9$&$2.47$\\
\hline
$\gamma_{E1M2}$&$1.19$&$0.23$&$ 0.0$ &$1.19$\\
\hline
$\gamma_{M1E2}$&$1.19$&$1.82$&$2.1$&$1.19$\\
\hline
$\gamma_{0}$&$4.74$&$-4.02$&$-0.7$&$-1.16$\\
\hline
$\gamma_{\pi}$&4.74&6.39&$9.3$&$6.14$\\
\hline
\end{tabular}
\end{center}} \hskip4.5cm
\parbox{3cm}{\vspace*{1.cm}
\caption{Proton spin polarizabilities obtained from different sources in units of $10^{-4}$ fm$^4$.}\label{table:polarizabilities}}
\end{table*}
Our results for the forward and backward spin polarizabilities are in agreement with
experimental values within error bars.

The results from strict  $\chi$PT in the second and third columns of
Table.~\ref{table:polarizabilities} follow from the formal results
presented in \cite{VijayaKumar:2000pv}  but using the parameter set of \cite{Gasparyan:2010xz}.
Comparing the $Q^3$ with the $Q^4$ results is significant since both results are determined
by the same parameters \cite{VijayaKumar:2000pv}. As can be seen from Table.~\ref{table:polarizabilities}
strict $\chi$PT appears not convergent for $\gamma_{E1E1}$ and $\gamma_{M1M1}$. In contrast for the off diagonal
elements $\gamma_{E1M2}$ and $\gamma_{M1E2}$ one may hope for slow convergence. This is interesting in view of the
large rescattering effects obtained for $\gamma_{E1E1}$ and $\gamma_{M1M1}$ in our approach. While for
$\gamma_{E1M2}$ and $\gamma_{M1E2}$ we recover the $\chi$PT results, for $\gamma_{E1E1}$ and $\gamma_{M1M1}$ we predict
significant modifications from the $\chi$PT results. It is amusing to see that our results for
$\gamma_{E1E1}$ and $\gamma_{M1M1}$ are quite compatible with the prediction of \cite{Holstein:1999uu} based
on fixed-$t$ unsubtracted DRs. However, note some discrepancies for $\gamma_{E1M2}$ and $\gamma_{M1E2}$.

The convergence pattern for $\gamma_{E1E1}$ and $\gamma_{M1M1}$ is reminiscent of the one for the
$s$-wave $\pi^0$ photoproduction multipoles \cite{Gasparyan:2010xz}. There are indications that
in our scheme there is significant accelerated convergence as compared to the $\chi$PT convergence.
In order to consolidate this expectation one may perform computations within our method to higher orders in the chiral 
Lagrangian.

\section{Spin polarization in Compton scattering}

The beam asymmetry $\Sigma_3$, beam-target asymmetries $\Sigma_{2z}$ and $\Sigma_{2x}$
for the proton Compton scattering  will be measured at MAMI \cite{MAMIA2-11/09} in a wide energy
range, starting from threshold. We refer to the Appendix~\ref{asymmetries} for their definitions. 

The main purpose of these measurements is the determination
of the spin polarizabilities of the proton. The appropriate tool to analyze the data set are
subtracted fixed-$t$ DRs \cite{Holstein:1999uu,Pasquini:2007hf}, which use as input empirical
pion photon-production amplitudes. In this approach the spin polarizabilities are free parameters that will be
adjusted to the data set. In order to shed further light on the intrinsic theoretical uncertainties it may be useful
to compare with our present approach based on the chiral Lagrangian.

\begin{figure}[t]
\centerline{
\includegraphics[height=6.5cm]{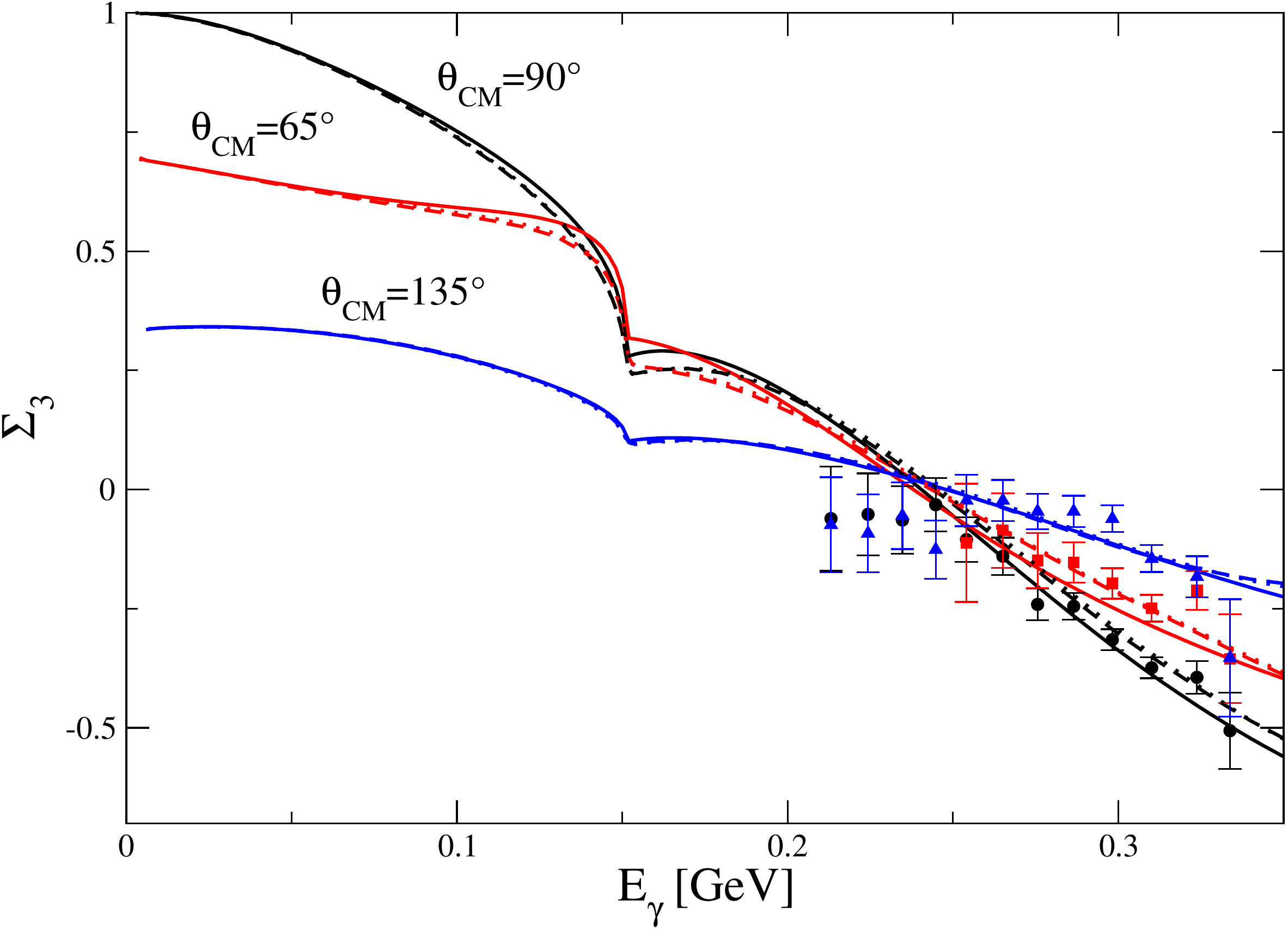}}
\caption{
The beam asymmetry $\Sigma_3$ for linearly polarized photon and unpolarized proton target,
as function of photon laboratory energy and for different center of mass scattering angles  as indicated.
The solid  curves result from our chiral approach. The dashed and dotted curves follow from subtracted DR calculation as explained in the text. The data are from \cite{Blanpied:2001ae} and correspond to the three scattering angles shown in the figure.}
\label{fig:Sigma3}
\end{figure}

We recall the definitions of the various spin observables in terms of helicity matrix elements, $\phi_i (\sqrt{s}\,,\theta)$,
and helicity partial-wave amplitudes, $ t^J_{\pm,ab}(\sqrt{s}\,)$, in the
convention used in \cite{Gasparyan:2010xz}. The  helicity matrix elements are expressed in terms of the
helicity partial-wave amplitudes as follows
\allowdisplaybreaks[1]
\begin{eqnarray}
\phi_1&=&\frac{m_p}{4\pi\,\sqrt{s}}\,\cos{\frac{\theta}2}\,\sum_J\left(t^J_{-,11}+t^J_{+,11}\right)
\big\{P'_{J+\frac{1}{2}}(\cos \theta) -P'_{J-\frac{1}{2}}(\cos \theta)\big\}\,,
\nonumber\\
\phi_2&=&-\frac{m_p}{4\pi\,\sqrt{s}}\,\sin{\frac{\theta}2}\,\sum_J \left(t^J_{-,11}-t^J_{+,11}\right)
\big\{P'_{J+\frac{1}{2}}(\cos \theta) +P'_{J-\frac{1}{2}}(\cos \theta)\big\}\,,
\nonumber\\
\phi_3&=&\frac{m_p}{4\pi\,\sqrt{s}}\,\frac{\sin\theta \,\sin{\frac{\theta}2}}{\sqrt{(J-\frac{1}{2})(J+\frac{3}{2})}}\,
\sum_J\left(t^J_{-,12}-t^J_{+,12}\right)
\big\{P''_{J+\frac{1}{2}}(\cos \theta)
\nonumber\\
&& \qquad \qquad \qquad \qquad  +\,P''_{J-\frac{1}{2}}(\cos \theta)\big\}\,,
\nonumber\\
\phi_4&=&\frac{m_p}{4\pi\,\sqrt{s}}\,\frac{\sin\theta \,\cos{\frac{\theta}2}}{\sqrt{(J-\frac{1}{2})(J+\frac{3}{2})}}\,
\sum_J\left(t^J_{-,12}+t^J_{+,12}\right)
\big\{P''_{J+\frac{1}{2}}(\cos \theta)
\nonumber\\
&& \qquad \qquad \qquad \qquad-\,P''_{J-\frac{1}{2}}(\cos \theta)\big\}\,,
\nonumber\\
\phi_5&=&\frac{m_p}{4\pi\,\sqrt{s}}\,\frac{2\,\cos^3{\frac{\theta}2}}{(J-\frac{1}{2})(J+\frac{3}{2})}\,
\sum_J\left(t^J_{-,22}+t^J_{+,22}\right)
\big\{-3\,(J-\textstyle\frac{1}{2})\,P''_{J-\frac{1}{2}}(\cos \theta)
\nonumber\\
&&\quad +\,(J-\textstyle\frac{1}{2})\,P''_{J+\frac{1}{2}}(\cos \theta) +2\,\big(P'''_{J-\frac{1}{2}}(\cos \theta)
-P'''_{J-\frac{3}{2}}(\cos \theta)\big)\big\}\,,
\nonumber\\
\phi_6&=&-\frac{m_p}{4\pi\,\sqrt{s}}\,\frac{2\sin^3{\frac{\theta}2}}{(J-\frac{1}{2})(J+\frac{3}{2})}\,
\sum_J\left(t^J_{-,22}-t^J_{+,22}\right)
\big\{3\,(J-\textstyle\frac{1}{2})\,P''_{J-\frac{1}{2}}(\cos \theta)
\nonumber\\
&& \quad +\,(J-\textstyle\frac{1}{2})\,P''_{J+\frac{1}{2}}(\cos \theta)
+2\,\big(P'''_{J-\frac{1}{2}}(\cos \theta) +P'''_{J-\frac{3}{2}}(\cos \theta)\big)\big\}\,,
\end{eqnarray}
with the scattering angle $\theta$ in the center of mass frame.
The spin observables have a transparent representation in terms of the helicity matrix elements
\begin{eqnarray}
 \frac{d\sigma}{d\Omega}&=&\frac{1}{2}\left(
|\phi_1|^2+|\phi_2|^2+2\,|\phi_3|^2+2\,|\phi_4|^2+|\phi_5|^2+|\phi_6|^2\right)\,,\nonumber\\
 \Sigma_3\,\frac{d\sigma}{d\Omega}&=&
\Re\,\big((\phi_1+\phi_5)\,\phi_3^*+(\phi_2-\phi_6)\,\phi_4^*\big)\,,\nonumber\\
\Sigma_{2x}\frac{d\sigma}{d\Omega}&=&\Re\big((\phi_5-\phi_1)\phi_4^*+(\phi_2+\phi_6)\phi_3^*\big)\, ,\nonumber\\
\Sigma_{2z}\frac{d\sigma}{d\Omega}&=&\frac{1}{2}\left(|\phi_5|^2+|\phi_6|^2-|\phi_1|^2-|\phi_2|^2\right)\,.
\end{eqnarray}
For the readers convenience we recall the relation of the helicity partial-wave amplitudes
with the covariant partial-wave amplitudes used in (\ref{polarizabilities:PW}). It holds
\begin{eqnarray}
T^J_{\pm,ab}(\sqrt{s}\,)&=& 2\,m_p\,
\left(\frac{s}{p^2_{\rm cm}}\right)^{J-\frac12}
\label{gn-trafo}\\
&  \times & \sum_{c,d}
\left(\begin{matrix}
\frac{\sqrt{s}}{p_{\rm cm}}&-\sqrt{\frac{2\,J-1}{2\,J+3}}\,\frac{m_p}{p_{\rm cm}}\\
0&1   \end{matrix}\right)_{ac}
t^J_{\pm,cd}(\sqrt{s}\,)
\left(\begin{matrix}
\frac{\sqrt{s}}{p_{\rm cm}}&0\\
-\sqrt{\frac{2\,J-1}{2\,J+3}}\,\frac{m_p}{p_{\rm cm}}&1   \end{matrix}\right)_{db}\,.
\nonumber
\end{eqnarray}

\begin{figure}[t]
\centerline{
\includegraphics*[height=6.5cm]{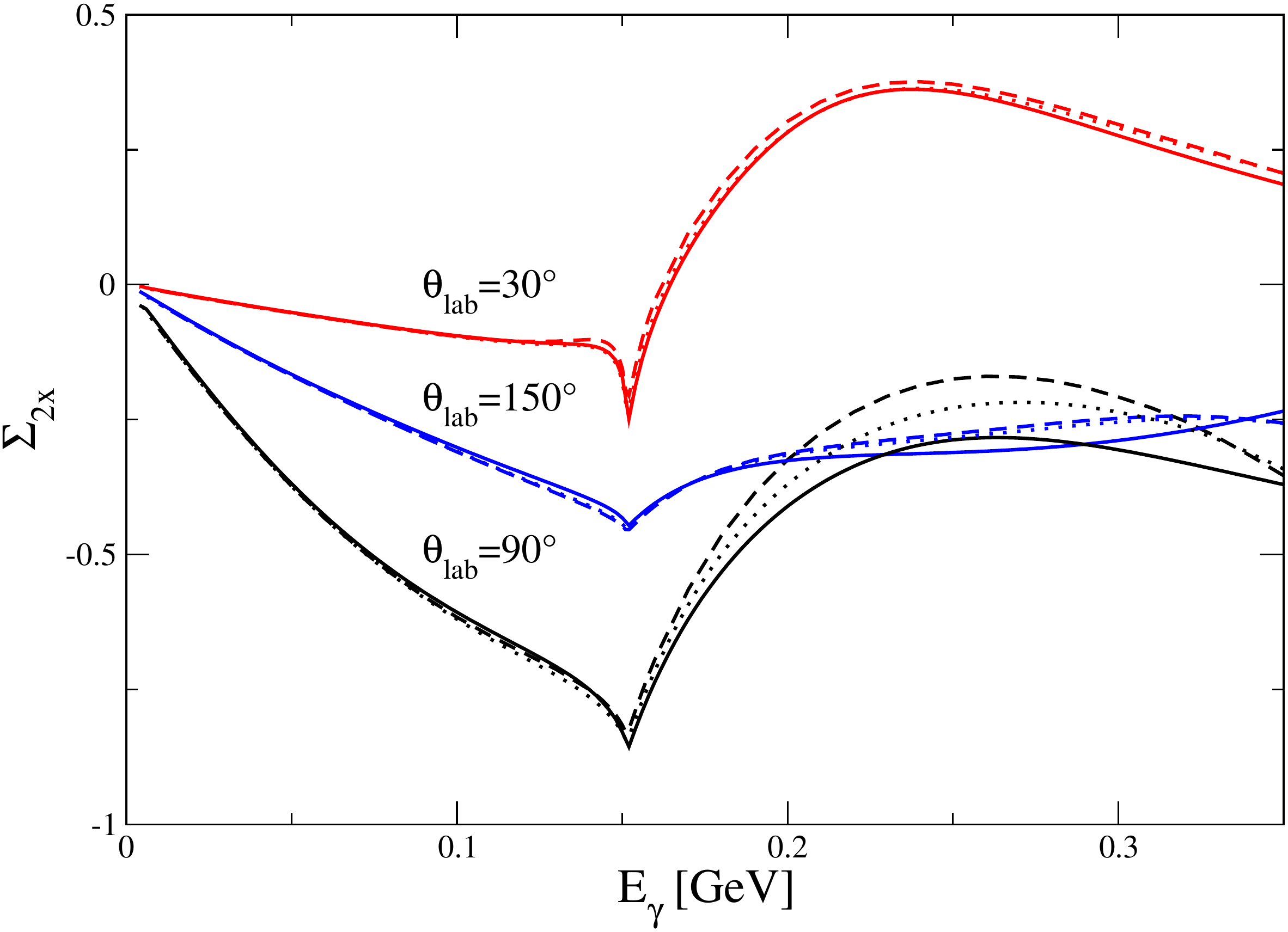}}
\caption{Double polarization asymmetry $\Sigma_{2x}$
for circularly polarized photon and target spin aligned transversely w.r.t. the photon momentum, as function of laboratory photon energy and scattering angles as indicated. The solid  curves result from our chiral approach. The dashed and dotted curves follow from  subtracted DR calculation as explained in the text. \label{fig:Sigma2x}}
\end{figure}
\begin{figure}[h]
\centerline{
\includegraphics*[width=8.5cm]{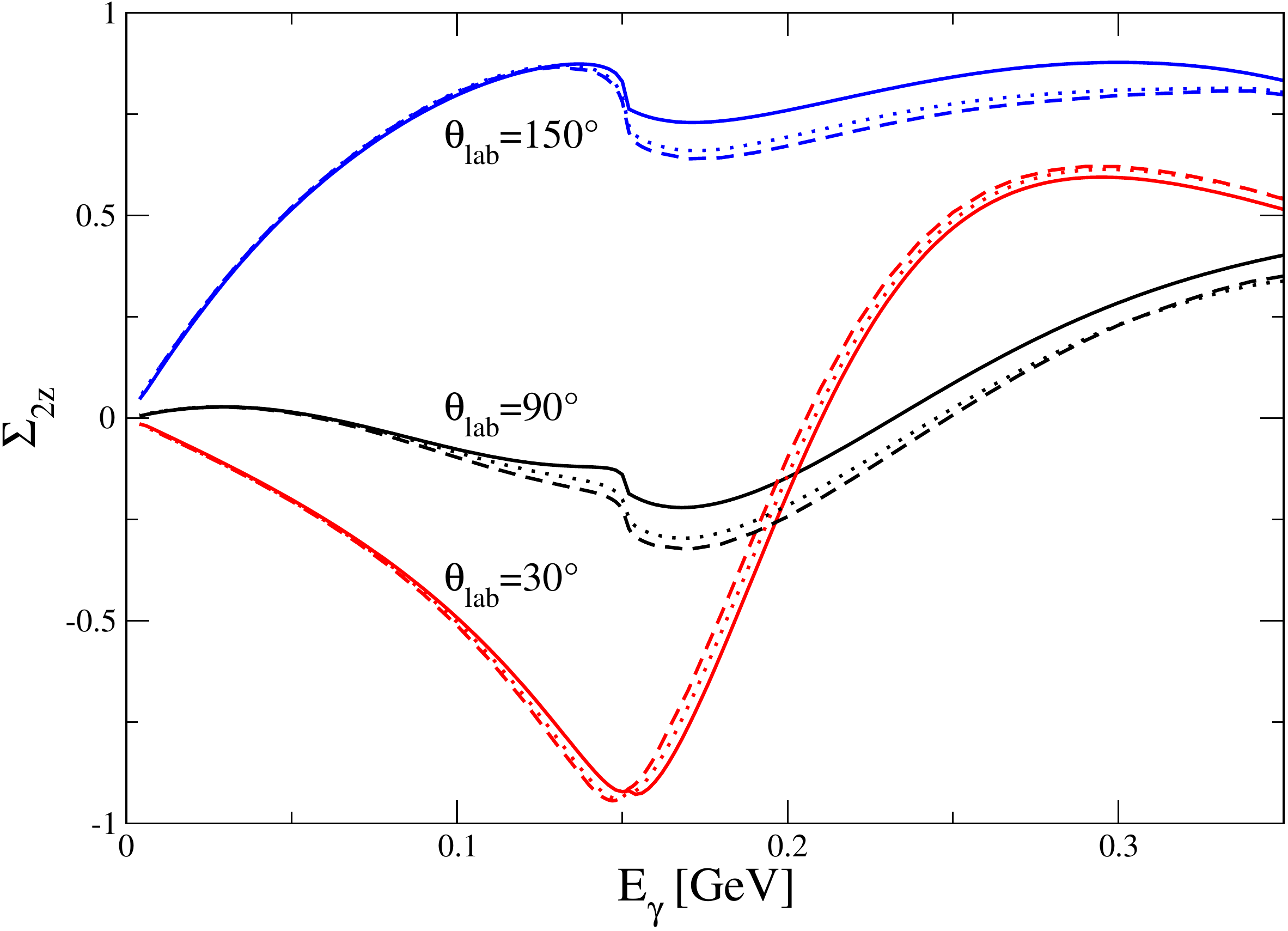}}
\caption{Double polarization asymmetry $\Sigma_{2z}$
for circularly polarized photon and target spin aligned longitudinally w.r.t. the photon momentum, as function of laboratory photon energy and scattering angles as indicated. The solid  curves result from our chiral approach. The dashed and dotted curves follow from subtracted DR calculation as explained in the text. \label{fig:Sigma2z}}
\end{figure}

Since in our approach there are no free parameters we compare our results (solid curves) in
Figs.~\ref{fig:Sigma3}-\ref{fig:Sigma2z} to calculations based on subtracted fixed-$t$  DRs (dashed and dotted curves).
For the dashed curves we use as  input our predictions for both the spin-dependent and the spin-independent polarizabilities. The dotted curves are obtained with the empirical spin-independent polarizabilities [22] together with the values of the spin polarizabilities given
in the second last column of Tab. 1. The difference between our predictions and the DR calculation
can provide a useful cross check and estimate residual uncertainties. For a discussion about possible sources of uncertainties in the DR calculation we refer to \cite{Pasquini:2007hf}.

In Fig. \ref{fig:Sigma3} the results for the beam asymmetry
are compared with available data \cite{Blanpied:2001ae} at different scattering angles.
New measurements at MAMI are expected to have a much better accuracy, typically at a few percent level.
In Figs.~\ref{fig:Sigma2x}-\ref{fig:Sigma2z} we show the results for the double polarization observables $\Sigma_{2x}$ and $\Sigma_{2z}$ which are mostly sensitive to the $\gamma_{E1E1}$ and $\gamma_{M1M1}$ polarizabilities, respectively.
The solid and dashed curves agree well below the pion production threshold and start deviating slightly
when the energy increases. The small difference between the two curves illustrates the precision one may expect for our results. Though the two methods are quite different in construction, they both rely on analyticity and unitarity which impose stringent constraints in the results for  Compton observables. The closeness of the dashed and dotted curves emphasize the similarity of the polarizabilities as predicted by our chiral and unsubtracted DR approach.

\section{Summary}

We studied nucleon Compton scattering from threshold up to
$E_{lab}=$ 350 MeV
within an approach developed in  \cite{Gasparyan:2010xz} and
based on an analytic extrapolation
of subthreshold amplitudes calculated in $\chi$PT. The Compton sector involves no additional free parameters.
Predictions for the spin polarizabilities were made within this scheme.
We found a nice agreement of the predicted values for the forward and backward spin polarizabilities
$\gamma_0$ and $\gamma_\pi$ with experimental values.  All four spin polarizabilities
$\gamma_{E1E1}$, $\gamma_{M1M1}$, $\gamma_{E1M2}$, $\gamma_{M1E2}$
are in a reasonable agreement with the calculations based on  fixed-$t$ DRs \cite{Holstein:1999uu,Pasquini:2007hf}. While for the transition polarizabilities
$\gamma_{E1M2}, \gamma_{M1E2}$ we recover the results of strict $\chi$PT, for the diagonal
$\gamma_{E1E1}, \gamma_{M1M1}$ elements we find significant effects from rescattering.

We also analyzed our predictions for several spin observables that will be measured at MAMI \cite{MAMIA2-11/09}.
Our results are close to the ones obtained from  fixed-$t$ DRs, although the two approaches are quite different in construction. This illustrates again the enormous predictive power of combining the unitarity and analyticity constraint in hadron physics. Furthermore, the comparison between the two calculations provides an important cross check and can be used to estimate the theoretical uncertainty  in calculations of Compton scattering observables.

\vskip1cm
{\bfseries{Acknowledgments}}
\vskip0.4cm
We thank M. Ostrick for stimulating discussions.

\appendix

\section{Appendix}
\label{asymmetries}
The beam asymmetry for photons which are linearly polarized either
parallel or perpendicular to the scattering plane and
unpolarized nucleons is defined as
\begin{eqnarray}
\Sigma_3  =\left(\frac{\sigma^\parallel - \sigma^\perp}
   {\sigma^\parallel + \sigma^\perp}\right),
  \end{eqnarray}
where $\sigma$ denotes the yield of particles corresponding to a particular polarization of the initial photon.

The asymmetries with circular photon
polarizations (left-hand or right-hand) are defined as
\begin{eqnarray}
 \Sigma_{2x}
   = \frac{\sigma^R_x-\sigma^L_x}{\sigma^R_x+\sigma^L_x}
      \, , \
  \Sigma_{2z}
   = \frac{\sigma^R_z-\sigma^L_z}{\sigma^R_z+\sigma^L_z}
   \end{eqnarray}
for the nucleon spin lying in the reaction
plane and aligned
in $\pm x$ or $\pm z$ directions, respectively. For more details we refer the reader to Ref.~\cite{Babusci:1998ww}.

\section{Appendix}
\label{PWamplitudes}
We decompose $F_{1-3}^\pm(\sqrt{s},t)$ in terms of partial wave amplitudes
\begin{eqnarray}
F_i^+(\sqrt{s},t)=\sum_{J,m,n} \,\frac{F_i^{mn}(\sqrt{s}\,)}{m_p^4} \,
\Big( \frac{\bar p_{\rm cm}\, p_{\rm cm}}{s} \Big)^{J+\frac{1}{2}-m-n}
P^{(m)}_{J+1/2-n}(\cos \theta )\,,
\end{eqnarray}
where we detail all non vanishing coefficients with
\allowdisplaybreaks[1]
\begin{eqnarray}
\frac{F^{20}_{1}}{m_p} &=&-\frac{(1-w) \left(1+2\,w-w^2\right) }{2 \,c_J\, w^4 }\,T^J_{+,21}
+ \frac{(1-w)^2 }{2\, (2 J+3) w^3}\,T^J_{-,22}
\nonumber\\
&+&\frac{\left(1+2\,w-w^2\right) }{2\,(2J+3)w^3}\,T^J_{+,22}
-\frac{(1-w)^2\,(1+w) }{2 \,c_J\, w^4 }\,T^J_{-,21}
\nonumber\\
\frac{F^{21}_{1}}{m_p} &=&\frac{(1-w)^4 (1+w)^3 }{8 \,c_J\, w^8} \,T^J_{+,21}
+\frac{(1-w)^3 (1+w)^2\,(1+3\,w)}{8\, (2 J+3)\,w^7}\, T^J_{+,22}
\nonumber\\
&+&\frac{(1-w)^3 (1+w)^2 \left(1+2\,w-w^2\right)}{8 \,c_J\, w^8}\, T^J_{-,21}
\nonumber\\
&+&\frac{(1-3\,w) (1-w)^2\,(1+w) \left(1+2\,w-w^2\right)}{8\, (2 J+3)\,w^7}\, T^J_{-,22}\,,
\nonumber\\
\frac{F^{31}_{1}}{m_p} &=&-\frac{(1-w)^5 (1+w)^3 \left(1+2\,w-w^2\right) }{8 \,c^2_J\, w^{11}}\,T^J_{+,22}
- \frac{(1-w^2)^5}{8\, c^2_J\, w^{11}}\, T^J_{-,22}\,,
\nonumber\\
\frac{F^{32}_{1}}{m_p} &=&\frac{(1-w^2)^7 }{32 \,c^2_J\, w^{15}}\,T^J_{+,22}
+ \frac{(1-w)^7 (1+w)^5 \left(1+2\,w-w^2\right) }{32 \,c^2_J\, w^{15}}\,T^J_{-,22}\,,
\end{eqnarray}
with  $c_J = \sqrt{(2\,J-1)\,(2\,J+3)}$ and $ w = \sqrt{s}/m_p$ and
\begin{eqnarray}
\frac{F^{10}_{2}}{m_p} &=&\sqrt{\frac{2 J-1}{2 J+3}}\, \frac{ 1-w}{2\,w^2}\, T^J_{+,21}
+\frac{1-2 J}{2\, (2J+3)\,w}\,T^J_{+,22} +\frac{1}{4 \,w^3}\,T^J_{-,11}
\nonumber\\
&-&\sqrt{\frac{2 J-1}{2 J+3}}\,\frac{1+w}{2\,w^2}\, T^J_{-,21}-\frac{1-2 J }{2\,(2J+3)\,w}\,T^J_{-,22}\,,
\nonumber\\
\frac{F^{11}_{2}}{m_p} &=& \frac{1-w^2}{8 \,c_J\, w^6}\left(6 J+1+(2 J+3)\,(w^2+w-1)\,w\right)T^J_{+,21}
-\frac{(1-w^2)^2 }{16\,w^7}\,T^J_{+,11}
\nonumber\\
&+&\frac{5-2 J+(2 J+3)\,(3\,w^3+4\,w^2-2\,w - 4)\,w}{8\, (2 J+3)\,w^5}\,T^J_{+,22}
\nonumber\\
&+&\sqrt{\frac{2 J+3}{2 J-1}}\,\frac{ (1-w)^3 (1+w)^2 }{8\, w^6}\,T^J_{-,21}
+\frac{(1-3\,w) (1-w)^2\,(1+w)}{8\, w^5}\, T^J_{-,22}\,,
\nonumber\\
\frac{F^{20}_{2}}{m_p} &=&\frac{-1-2\,w^3+3\,w^4  }{2 \,c_J\, w^6}\,T^J_{+,21}
+\frac{1-w^3 }{(2 J+3)\,w^4}\,T^J_{+,22}
\nonumber\\
&+&\frac{w^4-1}{2 \,c_J\, w^6}\, T^J_{-,21}
+\frac{1-2\,w^2+w^3 }{(2 J+3)\,w^4}\,T^J_{-,22}\,,
\nonumber\\
\frac{F^{21}_{2}}{m_p} &=&\frac{(1-w^2)^3 \left(1+w^2\right)}{8 \,c_J\, w^{10}} T^J_{+,21}
- \frac{(1-w^2)^2 }{8 \,c^2_J\, w^9} \Big(2 J+5-8 (J+1)\,w
\nonumber\\
&& \quad + 4 \,(1-2 J)\,w^2  + (2 J-7)\,(w-2)\,w^3\Big)\,T^J_{+,22}
\nonumber\\
&+&\frac{(1-w)^3 (1+w)^2 \left(1+w+w^2+3\,w^3\right)}{8 \,c_J\, w^{10}}\, T^J_{-,21}
\nonumber\\
&-&\frac{(1-w)^2\,(1+w) }{8 \,c^2_J\, w^9} \,\Big( 2 J+5-3 \,(2 J+1)\,w-4 \,(4 J+1)\,w^2
\nonumber\\
&& \quad + 2\,(7-2 J)\,w^3+9 \,(2 J+1)\,w^4+(7-2 J)\,w^5\Big)\,T^J_{-,22}\,,
\nonumber\\
\frac{F^{22}_{2}}{m_p} &=&\frac{(1-w^2)^4 }{32\,(2 J-1)\,w^{13}}\left(1-2\,w+2\,w^3+w^4\right) T^J_{+,22}
\nonumber\\
&+&\frac{(1-w)^7 (1+w)^5 }{32\,(2 J-1)\,w^{13}}\,T^J_{-,22}\,,
\nonumber\\
\frac{F^{30}_{2}}{m_p} &=&\frac{(1-w)^5 (1+w)^3 }{4 \,c^2_J\, w^9}\,T^J_{+,22}
+\frac{(1-w^2)^2  }{4 \,c^2_J\, w^9} \left(1-2\,w+2\,w^3+w^4\right)T^J_{-,22}\,,
\nonumber\\
\frac{F^{31}_{2}}{m_p} &=&-\frac{(1-w)^4 (1+w)^3}{8 \,c^2_J\, w^{13}}\left(1+w+2\,w^2+4\,w^3-w^4+w^5\right) T^J_{+,22}
\nonumber\\
&-&\frac{(1-w)^4 (1+w)^5 }{8 \,c^2_J\, w^{13}}  \left(1-w+3\,w^2-w^3\right) T^J_{-,22}\,,
\nonumber\\
\frac{F^{32}_{2}}{m_p} &=&\frac{(1-w)^6 (1+w)^7 }{64 \,c^2_J\, w^{17}} \left(1+w+3\,w^2-w^3\right)T^J_{+,22}
\nonumber\\
&+&\frac{(1-w)^6 (1+w)^5 }{64 \,c^2_J\, w^{17}}\left(1+3\,w+6 \,w^2+6\,w^3-5\, w^4+w^5\right)  T^J_{-,22}\,,
\end{eqnarray}
and
\begin{eqnarray}
F^{20}_{3} &=&\frac{2\,(1-w) }{\,c_J\, w^4}\,T^J_{+,21}-\frac{2 }{(2J+3) \,w^3}\,\Big( T^J_{+,22} - T^J_{-,22}\Big)
-\frac{2 }{\,c_J\, w^3}\,T^J_{-,21}\,,
\nonumber\\
F^{21}_{3} &=&\frac{(1-w^2)^2}{2 \,c_J\, w^7}\, T^J_{+,21}
-\frac{(1-w^2) \left(3-4\,w-3\,w^2\right) }{2\,(2 J+3)\,w^7}\,T^J_{+,22}
\nonumber\\
&-&\frac{(1-w)^3 (1+w)^2 }{2 \,c_J\, w^8}\,T^J_{-,21}
-\frac{(1-3\,w) (1-w)^2\,(1+w)}{2\,(2 J+3)\,w^7}\, T^J_{-,22}\,,
\nonumber\\
F^{31}_{3} &=&\frac{(1-w)^5 (1+w)^3 }{2 \,c^2_J\, w^{11}}\,T^J_{+,22}
+\frac{(1-w^2)^3 \left(1-2\,w-w^2\right)}{2 \,c^2_J\, w^{11}}\, T^J_{-,22}\,,
\nonumber\\
F^{32}_{3} &=&-\frac{(1-w^2)^5 \left(1-2\,w-w^2\right) }{8 \,c^2_J\, w^{15}}\,T^J_{+,22}
-\frac{(1-w)^7 (1+w)^5}{8 \,c^2_J\, w^{15}}\, T^J_{-,22}\,.
\end{eqnarray}
To obtain the expression for $F_i^-$ one can use the identities $F_i^+(-w, t)=F_i^-(w,t )$ and
(\ref{MacDowell-Compton}).
Note that $T^{J=1/2 }_{\pm,ab} =0$ for $a = 2$ or $b=2$.

\bibliography{1}
\bibliographystyle{elsart-num}

\end{document}